\documentclass[12pt]{article}
\usepackage{graphicx}
\usepackage{amssymb,amsmath,amsfonts,palatino,amsthm}
\usepackage{amssymb}
\usepackage{epstopdf}
\newcommand{\beq}{\begin{equation}}
\newcommand{\eeq}{\end{equation}}

\newcommand{\f}{\begin{equation}}
\newcommand{\ff}{\end{equation}}

\begin{document}

%%%%%%%%%%%%%%%%%%%%%%%%%%%%%%%%%%%%%%%%%%%%%%%%
\title{Prospects for constraining quantum gravity dispersion with near term observations}
\author{Giovanni Amelino-Camelia$^a$\thanks{amelino@roma1.infn.it} $\,$ and $$ Lee Smolin$^b$\thanks{lsmolin@perimeterinstitute.ca}
\\
\\
$^a$Dipartimento di Fisica, Universit\`a ``La Sapienza"\\
and Sez.~Roma1 INFN, P.le A. Moro 2, 00185 Roma, Italy\\
$^b$Perimeter Institute for Theoretical Physics,\\
31 Caroline Street North, Waterloo, Ontario N2J 2Y5, Canada}
\date{\today}
\maketitle

\begin{abstract}
We discuss the prospects for bounding and perhaps even measuring
quantum gravity effects on the dispersion of light using the
highest energy photons produced in gamma ray bursts measured by
the Fermi telescope. These prospects are brigher than might have
been expected as in the first 10 months of operation Fermi has
reported so far eight events with photons over $100 MeV$ seen by
its Large Area Telescope (LAT).   We review features of these
events which may bear on Planck scale phenomenology and
we discuss the possible implications for the alternative scenarios
for in-vacua dispersion coming from breaking or deforming of
Poincare invariance.  Among these are semi-conservative bounds,
which rely on some relatively weak assumptions about the sources,
on subluminal and superluminal in-vacuo dispersion.

We also propose that it may be possible to look for the arrival of
still higher energy photons and neutrinos from GRB's with energies
 in the range $10^{14}-10^{17} eV.$  In some cases the quantum
gravity dispersion effect would predict these arrivals to be
delayed or advanced by days to months from the GRB, giving a clean
separation of astrophysical source and spacetime propagation
effects.

\end{abstract}
\newpage

\tableofcontents
\newpage

\section{Introduction}

The possibility of probing the physics of quantum gravity with
high energy astrophysical observations has been discussed
seriously now for more than a
decade~\cite{aemn1}-\cite{leedsrPRL},
and there has been significant progress at Auger and other observatories,
but with the launch of the Fermi gamma ray
telescope~\cite{unoSCIENCE} in June 2008 it has become a reality.
This is because of the possibility of putting bounds on, or even
discovering, a generic consequence of quantum gravity models,
which is the dispersion of light governed by a scale\footnote{We
use units such that the Planck constant $\hbar$ and the
speed-of-light scale $c$ are set to $1$. Since we are considering
the possibility of in-vacuo dispersion we are implicitly assuming
 as operative definition
of the speed-of-light scale the value of the speed of light in the
infinite-wavelength limit.} $l_{QG}= \frac{1}{M_{QG}}$.  Here
$M_{QG}$ may be expected to be within  a few orders of magnitude
of $M_{Planck}= \frac{1}{\sqrt{G_N}}$.    This leads to a
variation in arrival time with energy,  roughly given by (see
later)
\begin{equation}
\Delta t  \simeq  \frac{\Delta E}{M_{QG}} L ~,
\label{delaySMALLz1}
\end{equation}
which could be as large as seconds to hours for photons in the
$GeV$ to $TeV$ range if the distance $L$ travelled is
cosmological.

Consequently, given the  timing accuracy of Fermi it has been
anticipated that after many events bounds could be put on $M_{QG}$
on the order of  $M_{Planck}$.  But, as we discuss here, the
situation is better than might have been hoped for because of
several features of the early data from the telescope, which
have so far been reported in papers, talks by collaboration
members, and notices for the Gamma-ray burst Coordinates Network
(GCN).

\begin{itemize}

\item{} There are already, in the first ten months of operation,
at least eight~\cite{unoSCIENCE,fermiERA,gwhen} Gamma Ray Bursts
(GRBs) detected whose spectrum extends to photons near or above
$1 \ GeV$ in energy, with the highest energy photon reported already at
$13 \ GeV$.

\item{}At least in one case (GRB080916C~\cite{unoSCIENCE})
the number of events at high energy was abundant enough to allow
spectral studies. And in several cases numerous high-energy
photons were observed.

\item{} This has allowed a remarkable early achievement, which is the
{\it raising of the conservative bound on $M_{QG}$, by an order of magnitude,
to within ten percent of the Planck mass,
based on the observation of a single GRB \cite{unoSCIENCE}!}

%counts falloff of the spectra with energy
%has the form of a shallow power
%law $N\sim E^{-\alpha}$
%with $\alpha < 2$, which gives more
%high energy photons than expected by some authors.

\item{} Some bursts are at  high redshift, with two
bursts with $z \approx 4$.

\item{} In these early events there are  clear trends that more
energetic photons arrive later, although the structure of the
events is complex.

\item{}  Already two short bursts have been observed at high
energies, which offer  approaches to
bounding in-vacuo dispersion complementary to those possible with long bursts.

\end{itemize}

The combination of these factors means that  even more stringent bounds
on $M_{QG}$ may be possible in the near future.  This also,
%bounds of $M_{QG} \approx M_{Planck}$ may
% be possible in the very near future. This,
as we will discuss, leads to a possibility of succeeding at the more difficult
challenge of measuring a nonzero $M_{QG}$ as data accumulates,
leading to a discovery of a breaking or deformation of special
relativity.  Making such a measurement is much harder than
putting a bound, because  the structure of the bursts are
complicated and there are astrophysical effects at the
sources over time scales comparable to $\Delta t$'s expected from
(\ref{delaySMALLz1}).  The challenge is then to find methodologies
which can be applied to the accumulated data sets which separate
astrophysical from possible quantum gravity effects.

To help prepare for facing this challenge we do the following in this paper.  First, in the next section
we survey the three basic
possible scenarios which lead to effects of (\ref{delaySMALLz1}) coming from either breaking or
deforming of lorentz invariance.  We also review the situation which obtained before Fermi to
discriminate amongst them observationally.

In section 3  we review and discuss some features of the GRB observations
reported so far by Fermi and explain the reasons for the
optimistic statements in the opening paragraphs. We review the
 reasoning behind the conservative bound on subluminal
propagation published so far\cite{unoSCIENCE} and propose new  sets of assumptions
that lead to new bounds, both on subliminal and superluminal
propagation.  These are somewhat less conservative than the bound published in
\cite{unoSCIENCE} but they may serve as sources of intuition for theoretical considerations.  We
also discuss comparisons between bounds obtained from Fermi
results and preliminary indications which had been previously
drawn from data on Mk501 and PKS2155-304.

In section 4 we discuss whether the data may eventually allow a
measurement of rather than a bound on $M_{QG}$.  We also
raise the possible role of new windows involving photons and/or
neutrinos at still higher energies  in the range of
$10^{14}$ to $10^{17} eV$.  These would be above the range that can
be seen from Fermi and would be observed by ground based
telescopes such as Auger and ICECUBE. To measure a quantum gravity effect
with these instruments would involve
correlations with GRBs with delays of days to months.  We will argue that
such observations are not impossible and would cleanly separate
astrophysical from quantum gravity effects.

Most of the literature on the phenomenology of in-vacuo dispersion concerns
models of dispersion with a single parameter, $M_{QG}$. In section
5 we discuss the possibility of models with two and more parameter
and discuss how they may be constrained by observations.

We close with conclusions.

\section{Options for one parameter modified dispersion relations}

The most basic question that can be asked about the quantum gravitational field, or
indeed of any physical system, is: {\it What is the symmetry of the ground-state?}

The ground state of general relativity is (ignoring the
cosmological constant) Minkowski spacetime, and its symmetry is
the Poincare group.  It is then natural to ask whether the
Poincare group is also the symmetry group of the quantum spacetime
geometry.  It may be, and this is {\it assumed} in several
approaches to quantum gravity, particularly perturbative
approaches such as perturbative general relativity and
perturbative string theory.  But it is natural to feel some
skepticism about the applicability of the Lorentz transformations
up to and beyond extreme cases where for example,  one
angstrom may be
Lorentz contracted by $25$ orders of magnitude to the Planck
length.  Experts will be aware that the intuitions of theorists on the ultimate fate
of Lorentz invariance are diverse, with accomplished theorists expressing views all along the
spectrum of expectations from the perfect validity to the complete breaking of
Lorentz transformations.  Our view is that the fate of Lorentz symmetry at the
extremes should be an experimental question and, happily, it is
becoming so.

Research   in quantum gravity phenomenology has focused on the
question of the fate of Lorentz invariance largely through the
lens of modifications of energy-momentum relations.  Over the last
years several scenarios have arisin for dispersion of light
motivated by theories and hypotheses about quantum gravity.  From
the perspective of experimental tests, these sort themselves into
three broad categories, which we will now discuss.  Note that as
we are discussing experimental tests we discuss them without
regard to our own views as theorists as to which, if any, is more
likely true. Similarly, we do not comment on whether it
appears to be more likely from a theoretical perspective that the
dispersion effects should first appear at linear or quadratic order in
$\frac{1}{M_{QG}}$.  We have the opportunity now with Fermi
to put strong bounds on the linear case so we focus on these
here.

\subsection{Lorentz symmetry breaking without effective field theory}

The first results on the implications of Planck-scale spacetime
structure for the persistence or not of the symmetries of special
relativity took the form~\cite{aemn1,gacPLB1997} of modifications
of the energy-momentum ``dispersion" relation
\begin{equation}
m^2 = E^2 -  {p}^2 +\Delta_{qg}(E, {p}^2;M_{QG}) ~,
\label{dispgeneral}
\end{equation}
where $E$ and $p$ denote energy and momentum of a particle of mass
$m$ and $M_{QG}$ is the reference/characteristic scale of
quantum-gravity effects, which is expected to be in some
relatively close neighborhood of the Planck scale. $\Delta_{qg}$ is a function with
dimensions $(mass)^2$.

In Refs.~\cite{aemn1,grbgac} it was observed that the
leading-order correction to the classical-spacetime dispersion
relation could be tested experimentally. We can parametrize these
leading-order correction in the ultrarelativistic ($E \gg m$)
limit as follows:
\begin{equation}
E \simeq p + \frac{{m}^2}{2p} - s_{{~}_{\! \! \! \! \pm}} \frac{1}{2} \frac{E^{\alpha+1}}{M_{QG}^\alpha}
~,
\label{dispAEMNSK}
\end{equation}
a parametrization which, in addition to $M_{QG}$, also involves the power $\alpha$,
expected to be an integer ($\alpha = 1$ for linear supppression by the quantum-gravity scale,
 $\alpha = 2$ for quadratic supppression by the quantum-gravity scale),
 and $s_{{~}_{\! \! \! \! \pm}} \in \{-1,1 \}$, which
 specifies the sign\footnote{This ``sign
parameter" $s_{{~}_{\! \! \! \! \pm}} = 1$ was denoted by $\xi$ in
Ref.~\cite{grbgac}. } of the correction $s_{{~}_{\! \! \! \!
\pm}} = 1$ gives the  ``subluminal" case whereby higher energy photons go slower, while $s_{{~}_{\! \! \!
\! \pm}} = - 1$ corresponds to the opposite ``superluminal" case.

 Starting with the studies reported in Refs.~\cite{kifune,kluz} the phenomenology based
 on the dispersion relation (\ref{dispAEMNSK}) also used the (unmodified) law of
 energy-momentum conservation, which in particular for
 a $a + b \rightarrow c + d$ particle-physics process gives
\begin{equation}
E_a + E_b = E_c + E_d
~,
\label{econs}
\end{equation}
\begin{equation}
 {p}_a +  {p}_b =  {p}_c +  {p}_d
~.
\label{pcons}
\end{equation}

Note that this framework breaks Lorentz symmetry, and it should therefore
be properly studied in a ``privileged" reference frame, such as the natural frame of CMB radiation.
We may call this
{\it naive Lorentz symmetry breaking} or NLSB\footnote{This framework emerged primarily
from the studies reported in Refs.~\cite{aemn1,gacPLB1997,grbgac,kifune,kluz}.}.

Moreover, it should be stressed that there need be no  dependence
of the correction terms on helicity/polarization and hence,  no
birefringence for the propagation of light\cite{aemn1,grbgac}. For
reasons that will be clear shortly, it turns out to be impossible
to describe the effects of this  scenario within the framework of
effective low-energy field theory in a classical spacetime. This
leads some theorists to be skeptical that this can be a realistic
framework.  Our view is that effective field theory can be an
important theoretical guide, but its ultimate validity is itself
an experimental question. Thus, theoretical expectations should
not be the basis of closing off experimental searches, especially
given the likelihood that new physical principles may come into
play at the Planck scale.

\subsection{Lorentz symmetry breaking within effective field theory}

Soon after the first papers on the NLSB scenario, Gambini and Pullin produced~\cite{gampul}
a first attempt of formalization within low-energy effective field theory.  This led to scenarios
we call lorentz-symmetry breaking in effective field theory, or (LSB-EFT).  They showed
that for correction terms that are only linearly suppressed by the Planck scale
($\alpha = 1$) one would inevitably end up predicting birefringence for light waves.

Note that while Gambini and Pullin worked within the framework of
loop quantum gravity~\cite{crLIVING,ashtNEW,thieREV,leeLQG}, their scenario depends on the assumption of
a particular and non-physical ground state for that theory.  Thus,
their scenario  should not be viewed as a definite prediction
of loop quantum gravity or more generally of other quantum
theories of gravity.

Unfortunately, this is typical of the current state of the art, in which
theories of  quantum gravity suggest possible new
phenomena that can be searched for experimentally,  without so far making
precise predictions for them~\cite{gacLRR}. At this phase
in our understanding it makes sense
to use semi-heuristic arguments based on the present understanding
of the various approaches to the quantum-gravity problem to derive
an intuition for the effects that could be expected, which are
then to be modeled phenomenologically.

For the details of modelling Planck-scale dispersion within an
effective-field-theory setup, a framework introduced by Myers and
Pospelov\cite{Myers2003} has proved very useful. It was shown
there, assuming essentially only that the lorentz symmetry
breaking effects are linear in $l_{Planck}$ and are characterized
by an external four-vector, that one arrives at a single possible
correction term for the Lagrangian density of electrodynamics:
\begin{equation}
 \mathcal L=-\frac{1}{4}F_{\mu\nu}F^{\mu\nu}
 +\frac{1}{2M_{Planck}} n^\alpha F_{\alpha\delta}n^\sigma
  \partial_\sigma(n_\beta\varepsilon^{\beta\delta\gamma\lambda}F_{\gamma\lambda})
 \label{eq:D5lagrangian}
\end{equation}
where  the four-vector $n_\alpha$ parameterizes the effect.

The (dimensionless) components of $n_\alpha$ of course take
different values in different reference frames, transforming
indeed as the components of a four-vector. A comprehensive
programme of investigations of this framework would
require~\cite{gacbiref} a phenomenology of the Myers-Pospelov
field theory exploring a four-dimensional parameter space,
$n_\alpha$, and contemplating the characteristic frame dependence
of the parameters $n_\alpha$. There is already a rather sizeable
literature on this phenomenology (see, {\it e.g.},
Refs.~\cite{gacLRR,gleiser,jlm,mattinLRR} and references therein)
but still fully focused on what turns out to be the simplest
possibility for the Myers-Pospelov framework, which is the one of
assuming to be in a reference frame where $n_\alpha$ only has a
time component, $n_\alpha = (n_0,0,0,0)$. Then, upon introducing
the conventional parametrization in terms of a dimensionless
parameter  $\xi \equiv (n_0)^3$, one can  rewrite
(\ref{eq:D5lagrangian}) as
\begin{equation}
 \mathcal L=-\frac{1}{4}F_{\mu\nu}F^{\mu\nu}+\frac{\xi}{2M_{Planck} }
  \varepsilon^{jkl} F_{0 j} \partial_0F_{k l}\, ,
 \label{eq:MP}
\end{equation}
and in particular one can exploit the simplifications provided
by spatial isotropy.

\subsection{Doubly special relativity}

The third option is doubly or deformed special
relativity~\cite{gacdsr,kowadsr,gacroxkowa,leedsrPRL,dsrnature,leedsrPRD,joaodsr}
which incorporates modifications or deformations of the Poincare
transformations without giving up on the principle of the
relativity of  inertial frames. The principle of the universality
of (the infrared limit of) the speed of light is joined by a
principle of the universality of a second, dimensional scale,
often taken to be the Planck energy.  This scenario can be
understood as the phenomenology arising from a quantum theory of
gravity in the limit, $\hbar \rightarrow 0$ and $G_{N} \rightarrow
0$ with the ratio $M_{Planck}= \sqrt{\frac{\hbar}{G_N}}$ held
fixed.

Over the last years it has been understood that this idea can be
expressed in several different frameworks and theories, leading to
a variety of phenomenologies. The most studied possibility is the
description in terms of ``Hopf algebras", a
generalization~\cite{lukie1,majidruegg} of the concept of Lie
algebra which appears to be relevant in some quantum pictures of
spacetime, such as spacetime
noncommutativity~\cite{gacdsr,kowadsr,jurekDSRnew}. Significant
progress has also been achieved by attempts to formulate DSR
theories in terms of an energy-dependent ``rainbow"
metric~\cite{rainbowDSR}.

At present DSR must be considered mainly a phenomenological
framework as it has not yet been fully incorporated into realistic
interacting quantum field theories. There have been several
heuristic arguments that DSR  follows from loop quantum
gravity\cite{kodadsr,lee-lqg-dsr1,lee-lqg-dsr2}, but no rigorous
proof. There are several results that indicate it is the case in
models of spacetime with $2+1$
dimensions~\cite{kodadsr,kodadsrkowa,kodadsrNEW2,kodadsrNEW}.
There are presently only partial results towards the construction
of interacting quantum field theories with DSR symmetry in $3+1$
dimensions. It is not known whether DSR can be realized in
string theory, although there is one positive result at the level
of the free bosonic string\cite{leeDSRstring}.

While all these results should be still considered
preliminary~\cite{dsrmg11dsr007}, the evidence available so far
encourages us to assume that a dispersion relation of the type
(\ref{dispAEMNSK}) could indeed be introduced in a DSR framework,
with deformed laws of transformation between observers but no
privileged class of observers (a ``deformation" of Poincar\'e
symmetry, but without breaking the symmetry).

For example, there has been a rather sizeable
literature~\cite{gacdsr,kowadsr,gacroxkowa} considering the
possibility that the dispersion relation be of the form
\begin{equation}
0= 8 M_{dsr}^2 \left[\cosh \left(\frac{E}{2 M_{dsr}}\right)
- \cosh \left(\frac{m}{2 M_{dsr}} \right) \right]
- {p}^2 \, e^{s_{{~}_{\! \! \! \! \pm}} \frac{E}{2 M_{dsr}} }
\label{dispKpoin}
\end{equation}
which for $E \ll M_{dsr}$ (of course also $M_{dsr}$ is expected to
have value close to the Planck scale) reproduces the dispersion
relation (\ref{dispAEMNSK}) with
 $\alpha=1$:
\begin{equation}
E \simeq p + \frac{{m}^2}{2p} - s_{{~}_{\! \! \! \! \pm}}
\frac{E^{2}}{2 M_{dsr}} ~. \label{dispDSR}
\end{equation}

Other forms of dispersion relations which have been explored
include\cite{leedsrPRD}
\begin{equation}
E^2 = \frac{p^2 }{(1+ s_\pm \frac{E}{4 M_{dsr} })^2} +m^2
\end{equation}
which also reduces at leading order to (\ref{dispDSR}).

Note that there are DSR scenarios which give either sign of
$s_\pm$, giving rise respectively to sub-luminal or superluminal
propagation.  But a given DSR scenario generally predicts a
parity even effect at leading order, so that one sign of $s_\pm$
holds for all photons, independent of polarization.  DSR
frameworks with quadratic ($\alpha =2$), rather than linear
($\alpha =1$), leading modification of the dispersion relation
have also been studied extensively (see, {\it e.g.},
Ref.~\cite{dsrIJMPrev} and references therein). It should also be
emphasized that there are special choices of DSR deformations
which leave the speed of light
unchanged\cite{jurekSPEED,unchanged}.

The consistency of a DSR framework requires two further
modifications of special relativistic physics that are not present
in either NLSB or LSB-EFT scenarios.  These are modifications
of the Poincare transformations connecting observations made by
different observers and modifications of the energy and momentum
conservation laws.  For example, the
 deformed laws of energy-momentum conservation, at leading
order in $\frac{1}{M_{dsr}}$ take the form\footnote{The exact form
of these deformed laws of energy-momentum conservation is a rather
messy combination of exponentials~\cite{dsrIJMPrev}, but in the
phenomenological applications we can foresee one only needs the
form at leading order in $\frac{1}{M_{dsr}}$.}
\begin{equation}
E_a + E_b - \frac{s_{{~}_{\! \! \! \! \pm}}}{2 M_{dsr}} p_a p_b
- E_c -E_d + \frac{s_{{~}_{\! \! \! \! \pm}}}{2 M_{dsr}} p_c p_d = 0
~,
\label{conservnewe}
\end{equation}
\begin{equation}
p_a + p_b - \frac{s_{{~}_{\! \! \! \! \pm}}}{2 M_{dsr}} (E_a p_b + E_b p_a)
- p_c -p_d + \frac{s_{{~}_{\! \! \! \! \pm}}}{2 M_{dsr}} (E_c p_d + E_d p_c) = 0
~.
\label{conservnewp}
\end{equation}
The deformations of transformation laws and energy-momentum conservation in DSR are
extensively discussed in the literature. As these play no role in time of arrival experiments, they
will not be further discussed here.

\subsection{In-vacuo dispersion in the NLSB, LSB-EFT and DSR frameworks}

From Eqs.~(\ref{dispAEMNSK}) and (\ref{dispDSR}) (respectively for
the NLSB and DSR frameworks) one easily
derives~\cite{aemn1,grbgac,gacdsr} that for two photons with
energy difference $\Delta E$ simultaneously emitted by a source at
relatively small redshift the times of arrival should differ by
\begin{equation}
\Delta t \big|_{small-z} \simeq  s_{{~}_{\! \! \! \! \pm}} \frac{
\Delta E}{M_{QG}} L ~, \label{delaySMALLz}
\end{equation}
where $L= H_0 z $ is distance of the source from Earth given in
terms of the Hubble expansion rate and the redshift. (In the DSR
framework $M_{QG} \approx M_{DSR}$.)
 For the ``subluminal" case, $s_{{~}_{\! \! \! \! \pm}}=1$, one has positive $\Delta t$
whenever $\Delta E$ is positive, meaning that for simulatenously emitted particles the one of lowest
energy is detected first. The opposite of course holds for the ``superluminal"
case , $s_{{~}_{\! \! \! \! \pm}}=-1$.

At large redshifts one should instead take into account the exact
(non-linear) dependence on redshift encoded in the
formula~\cite{piranZdependence,ellis07,ellis09}: The basic formula
for linear dependence \f \Delta t = \frac{\Delta E}{M_{QG}}
\frac{1}{H} \int_0^z dz \frac{1+z}{\sqrt{\Omega_\Lambda
+(1+z^3)\Omega_{Matter} }}
%\approx
%\frac{1}{25}\frac{M_{Pl}}{M_{QG}} \frac{\Delta E}{GeV} (z +
%\frac{z^2}{2} ) + ... \label{delayALLz}
\ff assuming $\Lambda$CDM cosmology with parameters
$\Omega_\Lambda$ and $\Omega_{Matter}$.

In the LSB-EFT framework there are similar effects of
energy-dependent speed of photons, but the effect carries opposite
sign for the two circular polarizations of light, {\it i.e.} it is
a birefringence effect.

\subsection{The situation before Fermi}

 There is a large literature~\cite{gacLRR,mattinLRR} on the phenomenology of lorentz symmetry
breaking, both naive and within effective field theory, and a
growing literature on the DSR phenomenology. Before Fermi, the bounds on the
in-vacuo dispersion expected in the NLSB or DSR contexts were
still two orders of magnitude below the Planck scale, even for the
case of Planck-scale-linear effects on which we are here focusing.
The best bound derived from $GRB$ data before Fermi was $M_{QG} >
2 \cdot 10^{17} GeV$ from~\cite{bestprior}.

However, in the case of
LSB-EFT, which has birefringent propagation, it has been established
that very stringent bounds can be derived from observations of
polarized radio galaxies. Assuming that the field-theoretic
LSB-EFT setup is spatially isotropic in the natural frame of the
CMB these bounds would exclude the entire range of values of $\xi$
that could be favoured from a quantum-gravity
perspective\footnote{If the actual quantum-gravity scale is within
a couple of orders of magnitude of the Planck one reaches a lowest
estimate of $\xi$ of about $10^{-2}$ while, assuming that the
field-theoretic LSB-EFT setup is spatially isotropic in the
natural frame of the CMB, even the most conservative
bounds~\cite{gleiser} would imply $\xi < 10^{-4}$.}. But, it has
been very recently noticed~\cite{gacbiref} that these bounds
become much weaker if one removes the assumption of isotropy in
the CMB frame, and that this assumption of isotropy is not
particularly preferred by the Myers-Pospelov setup. In light of
this, we shall in the following prudently consider the LSB-EFT
picture as still viable from a quantum-gravity perspective, but
perceive it as an approach that does not naturally match the
observations\footnote{We should also mention that the possibility
that quantum gravity dispersion competes with ordinary
electromagnetic dispersion in the intergalactic medium has been
considered~\cite{grbgac}, and it turns out that the latter is
negligible compared to the possibilities of the former in the
range of phenomena here of interest.}.

An important point is that so far as time of
flight experiments are concerned the NLSB and DSR scenarios predict the same
dispersion relations to leading order and hence make
the same leading order predictions. Thus,  to
distinguish them one must take into account experiments where
either or both of the modifications in transformation laws and
energy momentum conservation arise. As these are modified in DSR,
but not in NLSB, experiments that take them into account allow
the NLSB and DSR scenarios to be distinguished.

Observations where this is the case are tests of threshold effects
such as the GZK threshold predicted~\cite{kifune} for cosmic ray protons from
their scattering off the cosmological microwave background.
Similar predictions~\cite{kluz} hold for infrared photons scattering off of
the infrared background.  Because DSR maintains the principle of
relativity of inertial frames, the interactions involved can
always be evaluated in the centre of mass frame, where the
energies coming into the deformations from special relativity are
smaller.  Consequently, DSR makes, up to unobservably small
corrections~\cite{gacdsr}, the same predictions for threshold
experiments as ordinary special relativity.  However both lorentz
symmetry breaking scenarios, NLSB and LSB-EFT predict, for
suitable choices of parameters, order one modifications in the
positions of these thresholds.

To the extent that recent observations by Auger confirm the
standard special relativistic predictions for the GZK cutoff, the
lorentz symmetry breaking scenarios are disconfirmed, while the
DSR scenario remains unaffected. The only reservation that might be
made is that
it is possible to imagine that the parameters in the modified dispersion
relation are not universal.  The
GZK analysis applied to protons, and as this is the
only significant constraint for the NLSB scenario, it is possible to
hold open, as an experimental possibility, that the photon and
proton dispersion are governed by independent parameters.

However, in summary, it would be reasonable to say, with due
caution, that we entered the Fermi era with two strikes against the
LSB-EFT scenario, given by bifringence and GZK and one strike
against the NLSB scenario, given by GZK.

\section{First observations from Fermi relevant for quantum gravity phenomenology}

At present  there are reports~\cite{unoSCIENCE,fermiERA,gwhen} of $\sim 200$ GRBs observed at low
energies by Fermi's GBM, and for eight of these GRBs there are
reports of associated observations by Fermi's Large Area Telescope
(LAT), with photons with energies on the order of or greater than
$1 GeV$. With the exception of GRB080916C, which was throughly
described in Ref.~\cite{unoSCIENCE}, most of the information on
these bursts is presently only publicly available in resources,
such as notices for the Gamma-ray burst Coordinates Network (GCN),
that are not customarily in use in the quantum-gravity community,
which is part of the target readership of this paper. Hence, for
the convenience of theorists we summarize in Appendix A  the
information publicly
available~\cite{unoSCIENCE}-\cite{gwhen},\cite{z080916}-\cite{gcn090510}
on these 8 GRBs.   We list the information also in the following
table.

\begin{table}[htdp]
\begin{center}\begin{tabular}{c|c|c|c|c|c|c}GRB  & redshift & duration &
counts$|_{LAT}$& $E_{max}$ & $ t_i^{LAT}$ & $ t_f^{LAT}$
  \\\hline 080916C & 4.35 & long &strong & 13 GeV &
  4.5s& $>10^3 s$
   \\\hline 081024B & $~$ & short & $~$  & 3GeV & 0.2s & $~$
    \\\hline 090510 & 0.9 & short & strong  &  $>1$GeV  & $<1s$& $\gtrsim 60s$
     \\\hline 090328 & 0.7 & long & $~$  & $>1GeV$ & $~$&  $\approx 900s$
      \\\hline 090323 & 4 & long & strong  &  $>1$GeV  & $~$&
      $>10^3 s$
      \\\hline 090217 & $~$ & long & $~$  & $~$ & $\sim 1 s$& $\approx 20 s$
       \\\hline 080825C & $~$ & long & weak  & 0.6GeV & 3s&
       $>$40s
        \\\hline 081215A & $~$ & $~$ & weak  & 0.2GeV & $~$& $~$
\end{tabular}
\caption{GRBs seen by Fermi LAT with photon energies $\gtrsim
1GeV$.$ t_i^{LAT}$ is the time after the initial burst that high
energy photon seen by the LAT begin. $ t_f^{LAT}$ is the time
after the initial burst the high energy signals extend to.  For
references see the appendix. }
\end{center}
\label{GRBtable}
\end{table}

\subsection{Discussion of features of the bursts}

It is clear from the above table that there is a growing wealth of
information being gathered by Fermi which will be relevant for
testing the quantum gravity phenomenological scenarios we
discussed above.  It would be premature to draw rigorous
conclusions at this stage, before most of the data has been
analyzed and the results published by the Fermi collaboration.
Our aim here is not to compete with the work of observers, instead
we want only to draw attention to the potential inherent in what
is publicly known about the growing catalogue of events to resolve
a question at the heart of fundamental theoretical research.  To
this end we now briefly discuss some first conclusions which can
be drawn from the public  reports of these events.

\subsubsection{GRB080916C}

Among the 8 GRBs observed by the Fermi LAT GRB080916C is the one
that has generated most interest from a quantum-gravity
perspective. The feature on which these quantum-gravity discussion
have focused is the 13.2GeV photon detected by the LAT 16.5s after
the GBM trigger.    This allowed~\cite{unoSCIENCE}
 the Fermi collaboration to place a conservative bound on the parameter
$M_{QG}$ for pure dispersion in the ``subluminal" case (with
${s_{{~}_{\! \! \! \!  \pm}}}=1$). This is a remarkable achievement of the
Fermi collaboration, and we review it below.

There are other features that are relevant from the perspective of
dispersion studies and may eventually  prove even more valuable:
For GRB080916C Fermi detected $\sim 200$ high-energy ($>100 MeV$)
photons, allowing time-resolved spectral studies. And there was a
significant delay of $\simeq 4.5s$ between onset of $>100 MeV$ and
$\sim 100 keV$ radiation.  Also relevant  is the fact that the
time-resolved spectra for GRB080916C are well
fitted~\cite{unoSCIENCE,mesza0903} by an empirical
broken-power-law function (the so-called Band
function~\cite{band}) in the entire energy range, from $8 keV$ to
$\sim 10 GeV$, leading to the conjecture that a single emission
mechanism might have to describe what has been seen over this
broad range of energies. Moreover the $> 100 MeV$ emission lasts
at least $1400 s$, while photons with $ < 100 MeV$ are not
detected past 200 s. And it is for us particularly significant
that the time when the $> 100 MeV$ emission is detected ($\simeq
4.5s$ after the first $<5 MeV$ pulse) roughly coincides with the
onset of a second $<5 MeV$ pulse, but most of the emission in this
second ($[<5 MeV]\oplus[> 100 MeV]$) pulse shifts~\cite{mesza0905}
towards later times as higher energies are considered.

\subsubsection{GRB081024B and GRB090510}

Information that is somewhat complementary to the one provided by
GRB080916C could come from the two ``short" bursts in the sample,
which are GRB081024B and GRB090510. For GRB081024B  there was
no redshift determination  but preliminary reports
indicate~\cite{gwhen,gcn081024,longo081024} that the second peak
in GBM was seen $\simeq 0.2s$ after the (first-peak) trigger and a
few photons with energy $\gtrsim 100 MeV$ were observed in rough
coincidence with the second GBM peak, including a 3GeV photon. For
GRB090510, according to preliminary reports~\cite{gcn090510},
several multiGeV photons were detected within the first second of
the burst (whose inferred redshift is $\approx 0.9$).

This preliminary information on short GRBs is potentially very
significant for the outlook of studies of in-vacuo dispersion. And
it is to some extent unexpected~\cite{shortBUTGeV}, since before
these Fermi observations it had been argued that high energy
emission would be more likely for long GRBs. There are obvious
advantages for in-vacuo-dispersion studies when the analysis can
rely on sources of relatively short duration. And since it is
expected that the astrophysics of short GRBs is significantly
different~\cite{shortBUTGeV,tsviGRBrev,meszaSHORTlong} from the
one of long GRBs, the fact that both types of GRBs are well
observed at high energies could prove very valuable for efforts
aimed at disentangling propagation effects from effects at the
source.

 \subsubsection{Common features of the data}

In addition to the three we have just discussed the Fermi-LAT has
observed so far 5 other bursts. On some of these bursts  there
is still rather limited information, but preliminary reports
suggest that some features noticed in GRB080916C may be generic.
1) typically the onset of LAT events coincides with a second peak
in the GBM, a few seconds to fractions of a second, after the
first peak; 2)  high energy events last much longer than low
energy events; 3)  the number of LAT detections is often
relatively large.
%a relatively slow power law fall off in energy.

 \subsection{Constraints on subluminal in-vacuo dispersion}

Now we turn to conclusions that can be drawn from the data at this
early stage.   The first thing to mention for those interested in
possible  measurement of $M_{QG}$ is that the data cannot be
interpreted purely in terms of dispersion during travel.  In that
ideal situation there would be a simple linear relation between
photon arrival time and energy and that is not the case. On the
contrary it is typical that several lower-energy photons are
detected both before and after the detection of  the
highest-energy photon in the burst. There is also evidence that
the onset of arrival of higher energy photons comes in rough
coincidence with a second peak in low-energy detections, and the
presence of this feature for bursts at different   redshift may
(at least tentatively) encourage us to interpret it as  an
astrophysical effect.

The feature for which it appears most natural to invoke a role
played by quantum-gravity effects is the fact that, as most
clearly seen in GRB080916C (but supported also by the other
LAT-observed bursts), the second peak in the signal (first peak in
the LAT) shifts towards later times as higher energies are
considered.

But in any case, in light of these considerations, it is clear
that any extraction of a measurement of $M_{QG}$ requires some
methodology which models or averages out astrophysical effects.
This means that any fit to data relevant for measuring $M_{QG}$ is
likely to involve additional parameters, the minimum would be a
two parameter fit where the second parameter controls the
probability of emission of photons from the source as a function
of time since the burst and their energy (both in the frame of the
source).

It is far simpler to establish bounds on $M_{QG}$. We now turn to
this, focusing (as for most parts of this paper) on the case
$\alpha=1$ in which the effects depend linearly on (the inverse
of) the quantum-gravity scale.

 \subsubsection{Conservative bounds on subluminal propagation}

We begin with the ``subluminal case", with ${s_{{~}_{\! \! \! \!
 \pm}}}=1$. Here one can establish a lower bound
on $M_{QG}$ by simply measuring a distance to the source and a
delay time $\delta t$ for a certain high-energy photon. Assuming
that the photon left the source at the time of the initial burst
gives a value for $M_{QG}$.  But given that we cannot know it left
then, rather than a bit later, $\delta t$ is actually an upper
bound on the quantum gravity caused time delay, and hence the
corresponding $M_{QG}$ is a lower bound.

Using this methodology, the Fermi collaboration establishes a
bound on $M_{QG}$ using the $13.6 GeV$ photon of  GRB 080916C
which arrived $16 s$ after the initial burst~\cite{unoSCIENCE}
 \f
  M_{QG} > 1.3 \cdot 10^{18} GeV
 \ff
 {\it i.e.} roughly $M_{QG} > 0.1 M_P$.

 \subsubsection{Less conservative bounds on subluminal propagation using more structure in the data}

We note that this is counting time from the initial peak (the
``trigger" peak) of the burst. However, in light of the
observations we made above, it appears reasonable to assume that
at best quantum-gravity effects could have come into play in
generating a delay with respect to the time of the second
low-energy peak of GRB080916C (some $4.5 s$ after the first
low-energy peak), and this would then lead to a bound of
  \f
 M_{QG} > 1.8  \cdot 10^{18} GeV
 \ff
This of course cannot be considered as a conservative bound, but
we feel it is robust enough to be used tentatively as guidance for
further studies on the theory side (see below).

One might ask instead whether the delay in high energy photons arriving at the
second peak can itself be considered a result of in vacuo dispersion.  The problem is
that  the correspondence between the first peak of the LAT and the
second peak of the GBM (the low energy detector) is
particularly significant because the former is itself
a peak that receives contributions from a broad range
of energies.  Thus, if the delay of the first peak of the LAT and the first peak of the GBM
was due to dispersion then we should see even more dispersion
of the former than the "peak dispersion" between the first two peaks.
(Notice that $\Delta E$ between $100MeV$ and  $5 MeV$
is of course smaller than $\Delta E$ between $1GeV$ and $100MeV$)

We can illustrate this with a specific feature of GRB081024B.
In this case
preliminary analyses indicate~\cite{gwhen,gcn081024,longo081024}
there was a
small peak of photons with energy between $300$
and $500 MeV$
which arrived in coincidence with the second
low-energy peak, some
$0.2s$ after the first low-energy peak.
Then after another 0.2s a 3GeV photon arrived.
   Even without the redshift, which was not
measured in this case, we can use this to argue that it is
impossible that the delay between the first and second peak could
be a quantum-gravity effect. For any redshift, we can use  the $3
GeV$ photon to put a bound on $M_{QG}$. This would then imply that
any quantum-gravity delay acting on photons with a factor of ten
less energy could lead to a delay of no more than $0.04 s$.  Thus,
quantum gravity cannot account for the $0.2$ second delay between
the first low energy peak and the arrival of $300$ and $500 \ MeV$
photons coincident with the second low energy peak.

In the spirit of seeing what might be possible as the data
improves, we can ask what kind of bound on $M_{QG}$ would be
possible with a similar short burst, with the characteristics of
GRB081024B but with a measured redshift.  Suppose that the
redshift  of GRB081024B had been measured and found to be
$z_{081024B} \gtrsim 0.35$.  (This guess assumes that it is not
smaller than half of the smallest among the redshifts of the other
GRBs so far seen  by the LAT).   The result would have been a
bound of $M_{QG} \gtrsim 2.2  \cdot 10^{18} GeV$.

To see that this is not unreasonable, let us consider a second hypothetical argument of this
kind  based on the  preliminary information on GRB090510, which,
 has been announced~\cite{gcn090510} as a
burst  with several multiGeV photons within the first second after
the low-energy trigger. A bound of $M_{QG} \gtrsim 2.2  \cdot
10^{18} GeV$ would be confirmed by detecting a photon of, say, $2
GeV$ arriving from $z \approx 0.9$ within $0.4 s$ of the onset of
the LAT signal.

\subsubsection{Comparison with previous analyses of Mk501
and PKS2155-304}

Fermi has not been the only observatory making recent measurements
relevant for in vacuo dispersion.  The MAGIC and HESS detectors
have reported interesting observations of TeV flares from the AGNs
Mk501 and PKS2155-304, respectively. A study of spectral lags in
these observations  was found~\cite{ellis07,ellis09} to favour the
``subluminal" case (${s_{{~}_{\! \! \! \! \pm}}}=1$) with an
estimated measurement (rather than a bound) for $M_{QG}
=(0.98^{+0.77}_{-0.30}) \cdot 10^{18} GeV$. We may note that this
is  on the ``light side" of the range of values of $M_{QG}$ that
could be considered from a quantum-gravity perspective, and it
thus implies the effects of in vacuo dispersion are larger than
they would be for heavier $M_{QG}$, at or above $M_{Planck}$ . If
this estimate turns out to be correct, it is good news as it means
the discovery of quantum-gravity effects in Fermi's GRB data should
not be too challenging. In fact, with $M_{QG} \approx 10^{18} GeV$
for typical GRB redshifts of $\sim 1$, and for observations of
multi-GeV photons, the expected time delays would be of the order
of tens of seconds.  This time scale is safely larger than the
typical variability time scales one expects for the astrophysics
of GRBs.

It is then important to compare these measurements with the
results being reported from Fermi. The first thing to note is that
the conservative lower bound   $M_{QG} > 1.3 \cdot 10^{18} GeV$
established by the Fermi collaboration in Ref.~\cite{unoSCIENCE}
is  compatible within one standard deviation with the mentioned
estimate $M_{QG} =(0.98^{+0.77}_{-0.30}) \cdot 10^{18} GeV$ based
on previous observations of Mk501 and PKS2155-304. It is therefore
legitimate to continue to investigate this estimate.

On the other hand the observations we discussed above, concerning
the coincidence between the second peak of low-energy GRB signal
and the first peak of the $>100MeV$ GRB signal, appear to provide
encouragement for a somewhat higher value of  $M_{QG} $.   Our
``reasonably conservative" bound $M_{QG} > 1.8  \cdot 10^{18} GeV$
obtained from assuming the high energy photons started in
coincidence with the second peak of GRB080916C  is already more
than one standard deviation away from $M_{QG}
=(0.98^{+0.77}_{-0.30}) \cdot 10^{18} GeV$. And the remarks on
GRB082014B offered at the end of the previous subsection appear to
favour values of $M_{QG}$ that would be in significant
disagreement with the estimate $M_{QG}=(0.98^{+0.77}_{-0.30})
\cdot 10^{18} GeV$.

It would not be surprising if this disagreement between Fermi's
observations of GRBs and previous analyses of Mk501 and
PKS2155-304 was confirmed, since those results had been correctly
reported~\cite{ellis07,ellis09} as the outcome of ``conditional
analyses", relying on simplifying assumptions about the behaviour
of the sources. Still it is worth noticing that the
evidence of redshift dependence of the spectral lags
reported in Refs.~\cite{ellis07,ellis09} was uncovered by
considering average arrival times of particles in different energy
intervals, while both here and in Ref.~\cite{unoSCIENCE} the
analysis is focused on single photons and their specific detection
times. It is therefore  plausible that analyses of Fermi's
GRBs done in the same spirit of the ones previously applied to
Mk501 and PKS2155-304 ({\it i.e.} comparing average arrival times
of several photons detected in different energy intervals) might
uncover redshift-dependent effects consistent with the results
from  Mk501 and PKS2155-304,
 reported in Refs.~\cite{ellis07,ellis09}. We will
discuss in Section~5 their possible relevance for descriptions of
quantum-gravity effects on propagation that go beyond the
pure-dispersion picture.

\subsection{Bounds on superluminal propagation}

We now turn to discussion of putting possible bounds on
superliminal propagation, which is the case  ${s_{{~}_{\! \! \! \!
\pm}}}=-1$. This is important to do as from a theoretical
perspective there appears to be no compelling reason to prefer
either of the two possibilities, ${s_{{~}_{\! \! \! \! \pm}}}=1$
and ${s_{{~}_{\! \! \! \! \pm}}}=-1$. In the frameworks that are
at the basis of the NLSB and DSR pictures there is so far no
result  that may favour one or the other sign choice. Furthermore,
 the leading-order parity-violation effect
that arises in the LSB-EFT scenario scenario automatically provides both ``superluminal"
and ``subluminal" propagation, for the two circular polarizations
of photons.  Thus, in this case, one expects equal numbers of
subliminal and superluminal photons.

There are roughly two ways one might go about establishing bounds
on superluminal propagation: with photons that are detected and
with photons that are not detected.

\subsubsection{A bound from photons that are seen}

The first approach, using the photons that are detected, is
challenging because the high energy emissions are extended
in time. So while there is a clear signal for the beginning of a burst
from which retardation might be measured, there is not a clear
point from where advancement over lower energy photons might
be counted.

For example, on a first look at the data, particularly the data on
GRB080916C, one might naively deduce that it must be possible to
constrain the superluminal case rather tightly, since  the
data shows a  tendency of high-energy
particles to arrive, on average, later than low-energy ones. But
this feature  does not actually provide  evidence in
favour of subluminal propagation, since with our present, very
limited, understanding of the sources, it is possible that it be
fully of astrophysical origin. This  it also does not amount
to any evidence against superluminal propagation. The only safe
assumption on which one can anchor a conservative bound on
$M_{QG}$,  that all high-energy particles were not
emitted much before the low-energy particles that provide the GBM
trigger, is clearly useless from the point of view of establishing
a conservative bound on the case of superluminal propagation.

In the spirit of the type of considerations we offered in
Subsection~3.2.2, for the case of subluminal propagation, we can
look for arguments that allow to establish ``reasonably
conservative" (although not fully conservative) bounds on $M_{QG}$
for the case of superluminal propagation. Let us start by focusing
 our attention on
the first two photons with energy $\gtrsim 1 GeV$ that were
detected by the Fermi-LAT~\cite{unoSCIENCE} at $6.0 \pm 0.5$
seconds and at $7.0 \pm 0.5$ seconds after the trigger of
GRB080916C. It appears reasonably safe to assume that these
two photons were produced as part of a first main interval of
activity of the bursters which, from the data, we associate to the
time interval from the time of trigger to a time we
conservatively estimate to be $\lesssim 12$ seconds later
(see Fig.~1 of Ref.~\cite{unoSCIENCE}). On the basis of
these reasonably safe assumptions we deduce that a photon of
energy of at least 1 GeV after travelling a distance of z=4.3 had
not gained more than 5.5 seconds. From this we infer
 \f
  M_{QG}^{[{s_{{~}_{\! \! \! \!  \pm}}}=-1]} > 3.2 \cdot 10^{17} GeV
  ~.
  \label{superbound}
 \ff

One can arrive at  a comparable bound by considering the first
group of $>100MeV$ photons detected by LAT for GRB080916C. With
the  much higher total count one can clearly see~\cite{unoSCIENCE}
a reasonably smooth peak structure at $6.0\pm 0.5$ seconds after
the trigger, which (according to the observations on the structure
of the second peak of GRB080916C discussed above) we must place in
correspondence with a peak found at $5.3 \pm 0.7$ seconds after
the trigger for photons detected with energy between $260keV$ and
$5MeV$. On the basis of these observations we deduce that photons
with energy $\gtrsim 100MeV$ do not gain more than $ 0.5$ seconds
after travelling a distance of $z=4.3$, and in turn from this we
infer $M_{QG}^{[{s_{{~}_{\! \! \! \!  \pm}}}=-1]} \gtrsim 3.5
\cdot 10^{17} GeV$.

It is interesting that almost the same bound is obtained
from two independent  ``reasonably conservative" strategies, involving photons of different energies.
This is also, so far as we are aware, presently the best bound on superluminal propagation in the literature.
We may then suggest that (\ref{superbound}) be treated as a conservative upper
bound on the scale $M_{QG}^{[{s_{{~}_{\! \! \! \! \pm}}}=-1]}$ of possible
quantum-gravity-induced superluminal propagation (for the case of
effects that depend linearly on the inverse of such a scale).

Note that  we did not get beyond the level $\sim 3 \cdot 10^{17}
GeV$ because we could not exclude some sort of ``conspiracy" at
the source such that the observed delays of high-energy particles
be the result of even greater delays of emission at the source
which would be partly eroded along the way to the telescope.  For
example, if $M_{QG}^{[{s_{{~}_{\! \! \! \! \pm}}}=-1]} \approx 4
\cdot 10^{17} GeV$, which is a possibility not excluded by our
conservative bound, a 13.2 GeV photons arriving from $z\simeq 4.3$
should have gained along the way some 65 seconds, and as a result
it would have  needed some tuning to achieve an arrival time which
is 16s {\it after} the trigger, rather than some time before it.
While such conspiracies cannot be  excluded while attempting to
establish a robust bound, for the purpose of orienting our
theoretical intuitions we would argue that the data rather clearly
encourage us to focus future work on superluminal in-vacuo
dispersion on estimates that are significantly higher than $3
\cdot 10^{17} GeV$,  perhaps already in some neighborhood of
$M_{QG}^{[{s_{{~}_{\! \! \! \! \pm}}}=-1]} \sim M_{Planck}$.

\subsubsection{Implications of photons that are not seen}

A different kind of strategy, employing reasoning concerning
photons that are not seen, might be used to put stronger bounds,
particularly on the LSB-EFT scenario.  To do this one must assume
that there are no features of the source that would result in the
production of predominantly one helicity,  so we
expect equal numbers of subluminal and superluminal photons. Then,
if from a given source we see $N$ high energy photons within a
window in energy and time after the initial low energy burst, and
{\it no} photons in the same energy window within the same time before
the burst, one can set a limit on the probability that roughly $N$
photons of the opposite helicity which would be superluminal, were
produced but not detected.

To see how this might work, pick a candidate value of $\bar{M}_{QG}$ and consider a set of photons in a range of time and energy,
as follows. The photons must
arrive within a time $t \leq  t_0$ after the trigger with energies $E \geq E_0$ such that
the minimum time delay $\delta t =   \frac{ E_0}{\bar{M}_{QG}} L \geq t_0$.
These are chosen so that superluminal photons with the same characteristics should arrive
before the trigger.   Suppose that there are $N$ photons in this set.  Then there are roughly
$N$ missing photons, which should have arrived before the trigger if the LSB-EFT scenario is
true with that value of $\bar{M}_{QG}$, and the source does not emit predominantly in one helicity.

Let $p_{missed}$ be, for each photon that was observed, the probability that it might
have passed the detector and not been observed, and let $\bar{p}_{missed}$ be their average over
incident energy.  The probability that the  $N$ photons were missed is then their product, or
$p_{total}=p_{missed}^N$.
This is roughly the probability that the LSB-EFT scenario is correct with the chosen $\bar{M}_{QG}$
in spite of the fact that $N$
of the superluminal photons it predicts should have been detected were not.  That is, we can
say that with a probability $1-p_{total}$ that $M_{QG} > \bar{M}_{QG}$.

\section{Prospects for measuring a quantum gravity scale}

We now turn to the question of whether future experiments might
make possible a measurement of $M_{QG}$ rather than a bound.  As
we have discussed, this is much harder because of the possibility
of properties of the sources that mimic the effect of in-vacuo
dispersion, by introducing some correlation between the time of
emission and the energy of the particles. And, as we also stressed
above, the fact that the first results of the Fermi telescope do
not fit naturally within the most studied previous models of GRB
sources is likely to create a sort of competition between
postdiction of the observed features within accordingly taylored
astrophysical pictures and the possibility of in-vacuo-dispersion
effects.  If the quantum-gravity dispersion effects turned out to
be on the large side of the range of theory-favoured magnitudes,
as initially suggested by the preliminary analyses of AGNs
reported in Refs.~\cite{ellis07,ellis09} the competition with
model building on the astrophysics side might have been less
challenging, since with $M_{QG} \approx 10^{18} GeV$ for typical
GRB redshifts of $\sim 1$ and for observations of multi-GeV
photons the expected time delays would be of the order of tens of
seconds, a time scale that is safely larger than the typical
variability time scales one expects for the astrophysics of GRBs.
However, as we stressed in Subsection~3.2.3, the first
observations reported by Fermi, while still in principle
compatible with that estimate, provide the intuition that it is
likely that we should orient our speculations toward values of
$M_{QG}$ that are somewhat higher. This implies that the magnitude
of quantum-gravity effects may be comparable or even smaller than
the typical scales of time variability of GRBs.

It therefore appears that the best opportunities for discovering
 in-vacuo-dispersion effects will be based on  their
dependence on redshift. With correspondingly high statistics
(number of strong GRBs observed at different redshifts) it should
be possible to infer from analyses of this redshift dependence
some evidence of even small in-vacuo-dispersion effects.

In this section we would like to contemplate the
possibility of combining redshift-dependence analyses of Fermi
data with unusual events detected by
other observatories. In particular, we note that the
preliminary evidence of redshift-dependent effects found in
the analysis of Fermi data might acquire much greater significance
if some of the GRBs used in the analysis were also observed by
other telescopes, at energies higher than the ones accessible to
Fermi.

This possibility finds some encouragement in the first few
observations reported by Fermi. For GRB080916C there was positive
identification of $ \sim 200$ photons with energy $> 100 Mev$ and,
among these, 14 photons  with energy $>1 GeV$, which might suggest
that it is not unlikely that the signal is strong enough for
detection at even higher energies.
%a fall off not much worse than linear and
%certainly better than quadratic. One
%reaches similar conclusions from
%Fig.2A of Ref.~\cite{unoSCIENCE}, which can
%be illustrative of the ``count spectrum"
%of GRB080916C.

It is also encouraging that the  LAT signal tends to persist for
relatively long times after the trigger, in some cases of the
order of $10^3$ seconds.  This means that some telescopes that need
to be positioned in the direction of the burst (like MAGIC~\cite{magic}) should
have some chances of getting positioned in time.

Furthermore,  attempts to find VHE counterparts to Fermi-LAT GRBs are
particularly valuable for searches of superluminal effects, as we
shall stress rather forcefully in the last subsection.

\subsection{Photons of a few TeV}

The abundance of GeV photons detected by Fermi is encouraging
for the idea of observing some Fermi-LAT bursts also at TeV-photon
observatories. There is, however, an expectation (see, {\it
e.g.}, Ref.~\cite{piranIRABS} and references therein) of
significant absorption of $TeV$ photons due to electron-positron
pair production by IR background photons. However, our view is
that, nonetheless, these searches should be conducted without
reservations. In fact, the IR background is difficult to
determine, as direct measurements are problematic, owing to the
bright Galactic and Solar System foregrounds present. And in
recent years there have been several reports (see, {\it e.g.},
Ref.~\cite{piranIRABS} and references therein) of spectra of some
observed blazars that appear to be harder than anticipated,
considering the expected IR-background absorption.

Moreover, the NLSB
framework itself predicts a reduction of pair-production
absorption of $TeV$ photons~\cite{kluz,aus,aquilaPRD,gacpiranPRD,jacoTHRESH} (while
no such reduction is expected in DSR~\cite{gacdsr}), so this issue is mixed up
with that of in-vacuo dispersion. That is, if an NLSB
framework were true, there might be reduced absorption of TeV-scale
photons to be observed from GRBs.  This would be both indirect evidence
evidence for that scenario and permit the observation of a $TeV$ photons.

For $M_{QG} \sim M_{Planck}$ a 10TeV photon should acquire a delay
of about $10^3 s$ from z=4. This may be a manageable challenge for
some of the observatories which need to be directed toward the
burst.  A detection of such a photon identified with a $GRB$
would provide significant insight since $10^3 s$ is a time
scale that appears
 to be safely larger than the variability scales observed for
 these GRBs. The insight gained would still be very significant
 (though subject to more subtle analysis)
 for time delays of, say, $10 s$, as for example in the case of
 a $3TeV$ photon from z=1 in the case of $M_{QG} \sim 10
 M_{Planck}$. Delays of this magnitude however pose a challenge
 for observatories that need to be directed toward the burst.

\subsection{Photons with energies between $\sim 10^{14}$ and $\sim 10^{17}$ $eV$}

The abundance of GeV photons observed by Fermi may also give
indirect encouragement to the idea of detecting photons with
energy $\gtrsim 10^{14}eV$ from Fermi-LAT-observed bursts.
Moreover, while, as mentioned, some aspects of the spectral
analysis of GRB080916C and other bursts remain misterious from an
astrophysics perspective, some authors find in these data further
encouragement for the much-studied hypothesis that GRBs might be
responsible for (at least some of) the UHE cosmic rays. In turn
this would imply that GRBs are capable of producing UHE and VHE
photons ({\it e.g.} through decays of UHE neutral pions).

For VHE photons one would as well expect absorption by the soft
background photons, but for the same reasons mentioned in the
previous subsection we feel this should not necessarily discourage
such searches. In particular, we feel that such searches deserve
significant priority at the Auger cosmic-ray observatory~\cite{auger}.

For such high-energy photons the expected delays are
very large in the case of a linear quantum-gravity effect.  For example for $M_{QG} \sim
M_{Planck}$ a $10^{16}eV$ photon should acquire a delay of  $\sim
10^6 s$ ($\sim$ a month!) from z=4.
The possibility of identifying such a long delayed photon from a $GRB$
represents an extraordinary opportunity for
attempts to discover quantum-gravity dispersion.  But it also
pose observational challenges having to do with correctly attributing  such photons to a GRB
that had triggered much earlier.

 A key point is that even a single detection of
this kind could provide crucial input. We can envisage a stage,
possibly in the not too distant future, in which there are two
competing interpretations of the data on arrival times versus
energy of photons from GRBs , one from the quantum-gravity side
and one from the astrophysics side. A single detection with
such a huge time delay, but found to be in a time window
compatible with the magnitude of the effects predicted by the
quantum-gravity description of data at lower energies, could tilt
the balance in favour of that description.

\subsection{VHE neutrinos}

It has long been
recognized~\cite{mdrNEUTRI1,mdrNEUTRI2,mdrNEUTRI3,piranNatPhys,gacNatPhys}
that neutrinos can play a privileged role in the phenomenology of
the study of quantum-gravity effects on the propagation of
particles. This interest was centered mainly on the fact that
neutrinos appear to be our best chance, in the long run, to test for
dispersion effects suppressed quadratically by the
Planck scale. The
advantages of neutrinos from this perspective originate from the
fact that it gets easier to observe them from distant sources as
their energy increases, as a result of properties of the weak
interactions. And they travel essentially undisturbed by all
background fields in the universe
%, with no possible channels of absorption.

But it is also  possible that observatories such as ICECUBE~\cite{icecube}
could give decisive contributions to the present effort
of constraining or measuring quantum gravity effects suppressed
linearly by the Planck scale. For reasons that are completely
analogous to the ones discussed in the previous subsection for the
case of VHE photons, even a single such detection could play such a
decisive role, if it happened to corroborate an emerging
quantum-gravity interpretation of data at lower energies.

While the working assumption that GRBs produce VHE cosmic rays
leads us to expect that some VHE neutrinos are indeed produced by
gamma-ray bursters (through processes such as $p+\gamma
\rightarrow X+\pi^+ \rightarrow X + e^+ + \nu_e+ \nu_\mu +
{\bar{\nu}}_\mu$) all attempts of realistic estimates of
rates\cite{piranNatPhys}, also in relation to the sensitivities of
planned observatories, suggest that such searches of neutrinos
from GRBs might at best detect very few neutrinos. It is therefore
necessary to address concerns of a possible rejection of a genuine
detection of a neutrino from a LAT-observed GRB, which could be
missidentified as background/noise, especially if arriving with a
delay of, say, a month from the GRB trigger when, without the
quantum-gravity motivation, such detections would be completely
unexpected.

\subsection{Forward and backward in time}

We now turn from subluminal to superluminal propagation of very
high energy photons and neutrinos.  In the light of the
observations we reported in Section~2 and Subsection~3.3, searches
of VHE counterparts to LAT-observed GRBs are also very significant
for scenarios with superlulminal in-vacuo dispersion. And in this
respect it is worth stressing that, while, as discussed in
Subsection 3.3, placing bounds on superluminal effects is more
challenging than for subluminal effects, robust evidence for
superluminal propagation could be provided by simply establishing
that there are  some photons that arrive before the ones
composing the low-energy trigger.

While in Subsection 3.3 the desire to derive an absolutely
conservative bound led us to the prudent estimated bound of $
M_{QG}^{[{s_{{~}_{\! \! \! \!  \pm}}}=-1]} > 3.2 \cdot 10^{17}
GeV$, we shall here adopt as working assumption (for reasons which
we also discussed in Subsection 3.3) that our target should be at
the level of $M_{QG}^{[{s_{{~}_{\! \! \! \!  \pm}}}=-1]} \sim
M_{Planck}$, or even one or two orders of magnitude greater. For
such high values of the dispersion scale, and correspondingly
small magnitude of the dispersion effect, one would expect little
or no trace of it in data of the type of GRB080916C for energies
$\lesssim 10 GeV$. But dispersion effects of, say, 0.1 seconds for
$10GeV$ photons would imply dispersion effects of tens of seconds
for multi-TeV photons. This specific numerical estimate is
significant because the time interval between the first and the
second peak of GRB080916C is $\sim 4.5 s$, so in this scenario a
multi-TeV photon emitted together with the second peak of
GRB080916C would have been detected several seconds before the
low-energy GRB trigger. And in the same scenario a photon or
neutrino of, say, $10^{16}eV$ could have been detected $10^5 s$
before the GRB080916C trigger.

The only observation we know of that could provide encouragement
for these issues comes from the analysis reported in
Ref.~\cite{superlum910511}, which provided some (weak, 2.9
standard deviations) evidence of detection of photons with energy
$\sim \! 100 TeV$ from GRB910511 some 40 minutes before the
trigger of GRB910511. If observations of this sort were
established more robustly the interpretation would then be rather
straightforward. But it might be challenging to establish the
association with a GRB later seen at lower energies.

\section{Models of the quantum gravity vacuum with more than one parameter}

As we argued above  it appears natural to expect that a full
description of GRB data of the type of GRB080916C will require
quite several parameters, most (if not all) of which needed to model
the astrophyics of the system. Since such studies are in any case
necessary it is legitimate to contemplate the possibility of
uncovering scaling with redshift of more than one of the
parameters, and in particualr scaling that would not be consistent
with it parameterizing a property of the sources. It is therefore
of interest to discuss whether the quantum-gravity literature can
provide the basis for any positive expectations in this respect,
and in this section we want to comment briefly on this.

\subsection{Fuzzy dispersion}

The idea that quantum gravity would imply modified dispersion relations is
relatively new for quantum-gravity research; it started to be seriously discussed
 only in the second half of the 1990s.
 Before that discussions of possible effects of quantum
gravity on particle propagation mainly
concerned  stochastic or so-called fuzzy effects.   These were inspired by
speculations that quantum spacetime was ``foamy" so that spacetime structure would
 affect the average arrival time of a group
of particles, but would instead contribute to the spreading of
results of repeated measurements~\cite{fordFUZZY,perci2}.  One mechanism that
was proposed for this was that light cones would fluctuate in quantum gravity,
resulting in fluctuations in travel times of massless quanta.  There were also
studies of the idea that both dispersion and fuzziness could be
characteristic of the quantum-garvity vacuum.~\cite{iucaapap,uncpa1}.

To motivate this possibility, consider an event that produces
in a time $\Delta t^*$ a monochromatic burst of photons. Within
classical mechanics, and without in-vacuo dispersion, there is no
in-principle obstruction toward having such an event with
arbitrarily small $\Delta t^*$. But consider the effect of
turning on both quantum mechanics and in-vacuo dispersion of the
signal.

Quantum mechanics implies that the energy spread of a signal produced in a time
$\Delta t^*$ must necessarily be greater than $\hbar /\Delta t^*$, so
the burst cannot be sharply monochromatic if it is emitted in a
finite time interval $\Delta t^*$. Then
in-vacuo dispersion acts on the resulting $\Delta E$ to increase the time spread of the
signal seen at a distant telescope.

This conclusion is easily reached by describing the in-vacuo
dispersion  in terms of the formula
  \f
   v(E) = 1-\eta E/M_{Planck}
 \ff
and observing that, since $\Delta E \gtrsim \hbar/\Delta t^*$,  any
burst which at the source had duration $\Delta t^*$ should be
 characterized by a spreading of speeds of the particles that
 compose it given by
 \f
v(E)  \approx 1-\eta E/M_{Planck} \pm \eta \hbar  /(M_{Planck} \Delta
t^*) ~.
 \ff
As a result
observations of such a burst performed at a arge distances
$T$ from the source would not measure a spread of times of arrival
over an interval $\Delta t^*$, but rather
 \f
\Delta t_{meas}  \approx \eta T\hbar /(M_{Planck} \Delta t^*) ~,
 \ff
which can be much larger than the original $\Delta t^*$ if $T$ is
correspondingly large.

From a purely phenomenological point of view, one might then
contemplate an independent
contribution to fuzziness of the type~\cite{iucaapap,uncpa1}
 \f
v(E)  \simeq 1-\eta E/M_{Planck} \pm \eta /(M_{Planck} \Delta t^*)
\pm \eta_f E/M_{Planck} \label{fuzdis} \ff with $\eta_f$ a
phenomenological parameter to be determined experimentally but
expected to be within one or two orders of magnitude of 1.

One advantage to this kind of scenario is that, in contrast
to other lorentz symmetry breaking scenarios,  the GZK threshold remains essentially
 unaffected\cite{uncpa1}.

From a pure-phenomenology perspective a ``fuzzy dispersion"
of the type (\ref{fuzdis}) has some  characteristic
features. First, the predictions for the average arrival times
of a collection of particles within a particular energy
range are the same on average as in the pure-dispersion
picture. Thus, when it comes to the prediction of averaged
arrival times there is only one parameter, and it is the same
as the one parameter models.  This is significant because
the emission mechanisms are messy and likely introduce
randomness into the arrival times, thus the predictions of
quantum gravity models for
averaged arrival times with energy are more robust
than predictions for individual arrival times.

On top of these, the fuzzy picture introduces randomness
also in the quantum gravity predictions for arrival times
of individual photons.
This might make it possible to reconcile
observations that contradict each other under the
one parameter scenarios, and which also remain puzzling
after astrophysical sources of randomness are taken into account.
While this cannot be used to save scenarios that are
cleanly ruled out, it might become necessary if, for
 example, measurements based on averaged arrival times,
 using many particles, lead robustly to
measurements of values of $M_{QG}$ that are ruled out
by robust and conservative limits on $M_{QG}$ based
on arrival times of single photons (see
discussion in Subsection 3.2.3).

\subsection{Mixed parity dispersion}

The second possibility for a two parameter fit from quantum
gravity comes from the possibility that there is both an even and
an odd parity effect in dispersion, coming perhaps from a
fundamental chiral asymmetry in quantum gravity.  Indeed, a chiral
asymmetry is observed~\cite{chiral1,chiral2} in the formulation of loop quantum gravity,
and is parameterized by a parameter called the Immmirzi parameter.
Now, it has definitely not been shown that this leads to a mixed
parity dispersion of photon velocities but let us suppose it does.

Note that LSB-EFT predicts an odd parity effect in which
$
\delta v = - \beta   <s> \frac{E}{M_{QG}} $
where  $<s> $ is the
expectation value of chirality, a number which ranges $-1 \leq <s>
\leq 1 $, whereas NLSB and DSR predict an even parity effect
$ \delta v = - \alpha \frac{E}{M_{QG}} $, independent of helicity.
It is then possible to imagine that a quantum theory of gravity
might predict a mixed effect
 \f
\delta v = - (\alpha + \beta <s> ) \frac{E}{M_{QG}} \ff for
parameters $\alpha + \beta =1$.  To the extent that the
helicity of a photon can be treated as being essentially random in
GRB observations, this would induce a stochastic element in the
arrival times \f \delta t = (\alpha + \beta <s> ) \frac{E}{M_{QG}}
L \ff in the small $z$ approximation.

\section{Conclusions}

When, about a decade ago, the possibility of this type of studies
with Fermi (then known as GLAST) was first contemplated it
appeared that reaching Planck-scale sensitivity and beyond would be plausible
but challenging. It was reasonable to expect that
this might require a
large collection of GRB observations as well as
sophisticated methodologies for data analyses.

However, after less than a year of Fermi observations we already
have a robust bound at about $M_{QG} >0.1 M_P$, {\it from analysis
of a single $GRB$}.  The quality and quantity of data with
photons above $1 GeV$ makes it plausible that, with the large data
sets we will have after several years of Fermi observations, the
bounds may be pushed up to even a couple of orders of magnitude
beyond the Planck scale.  This would make it possible to fully
explore the range of values that could be favoured from a
quantum-gravity perspective, at least for a linear relation
between energy and velocity.

Even with the present data it may be possible to obtain bounds on
$M_{QG}$ that are significantly higher  than the present
conservative bound.  This is because of some unexpected features
of the GRB observations we briefly summarized above. One is the
coincidence between the arrival of the first multiGeV photons and
a second peak in the GBM. While this cannot be taken into account
when establishing conservative bounds, it is possible to suggest
that all the multiGeV photons originate at or after the second
peak, so that the delays should not be computed with respect to
the first GBM peak, but rather with respect to the second GBM
peak, thereby strengthening the bounds.  And in this respect
it is noteworthy that, according to preliminary reports, for
GRB081024B the LAT detected a 3GeV photon with a delay of only
$\approx 0.2 s$ with respect to the second GBM peak.  It is
unfortunate that there is no redshift in this case, but this
suggests a way in which a very good bound might be possible with a
short burst with known redshift.

To mention one scenario, it is entirely plausible that in a short
time there might be observed, say, 20GeV photon detected within,
say, $0.1 s$ of the first LAT peak for a GRB at a redshift of 4.5.
This could be used to establish a  bound at the level $ 4 \cdot
10^{20} GeV$  in the case of subluminal propagation. The fact that
placing bounds  (at least for ${s_{{~}_{\! \! \! \! \pm}}}=1$) is
relatively easy is manifest in the fact that such a powerful
result could be established even with the simple-minded strategies
of analysis that we discussed so far and even without any further
progress in the understanding of the astrophysics of GRBs.

 Having emphasized the bright prospect for
 setting bounds on $M_{QG}$, we then
 turned our attention to the greater challenge of discovering
quantum gravity effects by measuring a finite value of $M_{QG}$.
As we emphasized above, the prospects for this are more
challenging, in spite of there being several at least
superficially encouraging signs, such as the general feature of
time delays increasing with increasing energy.  Indeed, if we take
only the most energetic photons as data points, and assume there
are no astrophysical contributions to differential time delays, we
could imagine naively making a measurement of $M_{QG}$ within an
order of magnitude of $M_{Planck}$.  The problems, as we
emphasized above, are, first, that the data are not clean, so
there is no simple linear relation between arrival times and
energies for the high energy photons and second, because the time
scales for plausible astrophysical effects and the hypothesized
delays due to quantum gravity at these scales are comparable. It
seems then that a discovery of a quantum gravity time delay will
require a sophisticated methodology that deals with the
astrophysical contributions to the time delays either by
modeling them or by finding a way to subtract them out, also using
redshift-dependence analyses.

It would have been ideal if Fermi had confirmed the predictions of
one among the most studied emission mechanisms  in the
astrophysics literature. But we are in the opposite situation:
some aspects of GRB080916C are ``mysterious"~\cite{mesza0903} even
for some of the leading experts. With a more reliable reference to
a well-established astrophysical picture the discovery even of
particularly small effects (such as in cases in which $M_{QG} \sim
M_{Planck}$ or even one or two orders of magnitude bigger) could
be achieved with relatively small samples of GRBs at different
redshifts. In that ideal scenario both the redshift dependence and
the comparison to the expectations of the reference astrophysical
model could have been used in such searches. But already with
these first few Fermi-LAT observations it is rather clear that
attempts to make a discovery of a quantum gravity effect will have
to be conducted in conditions that are significantly different
from this ideal scenario.

Thus, it is possible that in the not too distant future we will be faced with a situation in which
there is a competition and perhaps even a degeneracy between astrophysical and quantum gravity explanations of time delays
seen in GRBs.   It may very well be that Lorentz symmetry is not broken at linear order in $l_{Planck}$
so that astrophysical explanations suffice to explain the data from GRBs.  But, given what is at stake
for fundamental physics, it would be foolish to assume this while inventing astrophysical explanations for
the time delays in the data, thus risking hiding what could be a fundamental experimental discovery of a
breakdown or modification of special relativity theory.

It is then very important to search for ways to break this competition or degeneracy.  To do this we
 turned in section 4 to the prospect of observing photons and neutrinos
at higher energies above the range of Fermi's LAT, up to $10^{17} ev$.  The quantum
gravity  time delays in these cases would be hours to months, so there would be a clean
separation of astrophysical and quantum gravity time scales.  The prospect of making such
observations is challenging, but we argued that the results would be very important.  Moreoever,
as we emphasized above, the arguments sometimes given for not searching for $TeV$ scale
photons from cosmological distances, because of absorption by the infrared background, cannot
be relied on as it rests on assumptions about the applicability of special relativity which are being
tested here.  Indeed, the NLSB scenario predicts that the threshold for that absorption can be
moved significantly to allow $TeV$ photons to reach us from cosmological distances.

We close with messages to both observers and theorists.  To
observers we would emphasize the opportunity for putting very
significant bounds on some or all of the quantum gravity scenarios
for modifying or breaking Lorentz invariance.  We would also
emphasize the importance of experiments and analyses that could
lift the degeneracy between astrophysical and quantum gravity
explanations for a correlation between photon energy and delay in
arrival times after initial bursts of GRBs.   To quantum gravity
theorists we suggest urgent attention be given to any possibility
of deriving predictions for these observations from theories of
quantum gravity, otherwise it may be only a matter of months to a
year or two before we theorists are demoted to the role of
postdictors of great experimental discoveries.

\section*{ACKNOWLEDGEMENTS}

We are thankful to Antonio Capone, Sabine Hossenfelder and Flavio
Mercati for discussion during the preparation of this paper. We
are also grateful to Camilla and Edoardo Amelino-Camelia and Kai
Smolin for inspiration, particularly regarding  the superluminal
case.  Research at Perimeter Institute for Theoretical Physics is
supported in part by the Government of Canada through NSERC and by
the Province of Ontario through MRI. G.~A.-C. is supported in part
by grant RFP2-08-02 from The Foundational Questions Institute
(fqxi.org).

\appendix.

\section{Appendix: The growing list of relevant GRB's with GeV scale photons}

For the convenience of theorists we summarize here the publicly
available information on the GRBs discussed above.

\begin{enumerate}

\item{}{\bf GRB 080916C}: We described in some detail  this very
strong long burst in Section~3. Photons were
detected~\cite{unoSCIENCE} by LAT up to $\sim 13 GeV$ (three
photons above $6GeV$) and the overall strength of the LAT signal
was such that time-resolved spectral studies could be
performed~\cite{unoSCIENCE}. Afterglow studies~\cite{z080916}
allowed to determine a redshift of $4.35\pm 0.3$.

\item{}{\bf GRB081024B}: This was the first short burst (described
in Refs.~\cite{unoSCIENCE,gcn081024,longo081024}) with signal
above $1 GeV$ (with maximum energy of $3GeV$), generating also
some puzzlement~\cite{shortBUTGeV} with respect to the
characterization of short bursts that were fashionable before the
Fermi era.

\item{}{\bf GRB090510}:  For this short burst (described in
Ref.~\cite{gcn090510}), at a redshift of $\approx 0.9$, the Fermi
LAT  detected more than 50 events above 100 MeV ( at least 10
above 1 GeV) within 1 second of the low-energy trigger and more
than 150 events above 100 MeV (at least 20 above 1 GeV) in the
first minute after the trigger.

\item{}{\bf GRB090328}:  In this burst (described in
Ref.~\cite{gcn090328}) the emission in the LAT lasts up around
$900s$ after the trigger, with the highest energy events (some
with $>1GeV$) arriving houndreds of seconds late. Afterglow
studies~\cite{z090328} allowed to determine a redshift of $0.7$.

\item{}{\bf GRB090323}: In this burst (described in
Ref.~\cite{gcn090323}) the emission is observed in the LAT up to a
few GeV, starting a few seconds after the GBM trigger time, and
lasting $\sim 2 \cdot 10^3 s$. Afterglow studies~\cite{z090323}
allowed to determine a redshift of $\sim 4$.

\item{}{\bf GRB090217}: In this burst (described in
Ref.~\cite{gcn090217}) the high-energy emission commences several
seconds after the GBM trigger and continues for up to 20 seconds
after the GBM trigger.

\item{}{\bf GRB 080825C}: This was the first GRB seen by the Fermi
LAT~\cite{unoSCIENCE}. The LAT signal~\cite{gcn080825} was
composed only of photons with energies below $1GeV$. Even though
the signal in the LAT is rather weak~\cite{gcn080825} it provides
significant evidence that the high-energy component has a delayed
onset~\cite{unoSCIENCE} and persistence up to 35s after the
trigger.

\item{}{\bf  GRB 081215A}: This burst (described in
Ref.~\cite{gcn081215a})was at a large angle to the LAT boresight,
and as a result
 neither directional nor energy information could be obtained
with the standard analysis procedures. A preliminary analysis however provides
evidence~\cite{gcn081215a}
of over 100 events above background, with energy presumably $\lesssim 200 MeV$,
detected
within a 0.5 s interval in coincidence with the main GBM
peak.

\end{enumerate}

\end{document}